\documentclass{sigchi}

\CopyrightYear{2018}
\setcopyright{acmlicensed}
\doi{https://doi.org/10.1145/3172944.3172957}
\isbn{978-1-4503-4945-1/18/03}
\conferenceinfo{IUI'18,}{March 7--11, 2018, Tokyo, Japan}
\acmPrice{\$15.00}

\toappear{\scriptsize Permission to make digital or hard copies of all or part of this work for personal or classroom use is granted without fee provided that copies are not made or distributed for profit or commercial advantage and that copies bear this notice and the full citation on the first page. Copyrights for components of this work owned by others than ACM must be honored. Abstracting with credit is permitted. To copy otherwise, or republish, to post on servers or to redistribute to lists, requires prior specific permission and/or a fee. Request permissions from permissions@acm.org. \\
{\emph{IUI 2018, March 7--11, 2018, Tokyo, Japan.} } \\
Copyright is held by the owner/author(s). Publication rights licensed to ACM. \\
ACM ISBN 978-1-4503-4945-1/18/03\ ...\$15.00.\\
https://doi.org/10.1145/3172944.3172957}

\usepackage{balance}       
\usepackage{graphics}      
\usepackage[T1]{fontenc}   
\usepackage{txfonts}
\usepackage{mathptmx}
\usepackage[pdflang={en-US},pdftex]{hyperref}
\usepackage{color}
\usepackage{booktabs}
\usepackage{textcomp}
\usepackage{makecell}
\usepackage{mathtools}
\usepackage{amsmath}
\usepackage{multirow}
\usepackage{soul}

\usepackage{microtype}        
\usepackage{ccicons}          

\usepackage{todonotes}

\def\plaintitle{Session-based Suggestion of Topics for Geographic Exploratory Search}

\def\emptyauthor{}
\def\plainkeywords{Geographical Information Retrieval; Session-based Concept Suggestion; Query Expansion.}

\makeatletter
\def\url@leostyle{%
  \@ifundefined{selectfont}{
    \def\UrlFont{\sf}
  }{
    \def\UrlFont{\small\bf\ttfamily}
  }}
\makeatother
\def\pprw{8.5in}
\def\pprh{11in}

\definecolor{linkColor}{RGB}{6,125,233}
\hypersetup{%
  pdftitle={\plaintitle},
  pdfauthor={\emptyauthor},
  pdfkeywords={\plainkeywords},
  pdfdisplaydoctitle=true, 
  bookmarksnumbered,
  pdfstartview={FitH},
  colorlinks,
  citecolor=black,
  filecolor=black,
  linkcolor=black,
  urlcolor=linkColor,
  breaklinks=true,
  hypertexnames=false
}

\begin{document}

\title{\plaintitle}

\numberofauthors{2}
\author{%
  \alignauthor{Noemi Mauro\\
    \affaddr{Computer Science Dept., University of Turin}\\
    \affaddr{Torino, Italy}\\
    \email{noemi.mauro@unito.it}}\\
  \alignauthor{Liliana Ardissono\\
    \affaddr{Computer Science Dept., University of Turin}\\
    \affaddr{Torino, Italy}\\
    \email{liliana.ardissono@unito.it}}
}

\maketitle

\begin{abstract}
Exploratory information search can challenge users in the formulation of efficacious search queries. Moreover, complex information spaces, such as those managed by Geographical Information Systems, can disorient people, making it difficult to find relevant data. In order to address these issues, we developed a session-based suggestion model that proposes concepts as a {\em ``you might also be interested in''} function, by taking the user's previous queries into account. Our model can be applied to incrementally generate suggestions in interactive search. It can be used for query expansion, and in general to guide users in the exploration of possibly complex spaces of data categories.

Our model is based on a concept co-occurrence graph that describes how frequently concepts are searched together in search sessions. Starting from an ontological domain representation, we generated the graph by analyzing the query log of a major search engine. Moreover, we identified clusters of ontology concepts which frequently co-occur in the sessions of the log via community detection on the graph.
The evaluation of our model provided satisfactory accuracy results.
\end{abstract}



\section{Introduction}
\label{introduction}

During an information search task, various issues challenge the specification of efficacious queries. Firstly, as discussed by Belkin in \cite{Belkin:80}, users are often unable to use the appropriate keywords because they are looking for data they do not know about. Secondly, complex information spaces, organized in several topic categories, can overload and disorient people, preventing them from finding the information they need; e.g., see \cite{Ratzan:04}. 
In order to address these issues, several systems present lists, or hierarchies, of information categories to search for. However, in this way they expose users to possibly large sets of options to choose from. 

Our work focuses on exploratory search, which is affected by the above issues because the users' information goals are ill-defined; see \cite{Marchiorini:06,White-Roth:06}. We aim at helping users orientate themselves by suggesting relevant topics (concepts) to be explored, given their observed search behavior. More specifically, we aim at guiding data exploration by proposing a small set of concepts that the user might be interested in, given the search context, as a {\em ``you might also be interested in''} function that helps her/him complete the search.
We adopt an associative information retrieval model \cite{Giuliano-Jones:62}:
our hypothesis is that, by analyzing the first query(ies) of a search session, and by taking into account which types of data are often searched together by people, the system can help the user identify further topics he or she might be interested in, by suggesting terms for query expansion. E.g., if the user looks for kindergartens in a town, he or she might also be interested in play and sports areas, as well as in other data related with children activities. Our work aims at supporting the suggestion of such concepts.
Specifically, we aim at answering the following research question:

{\em H1: Can the data about the concepts frequently searched together by people within a search session be exploited to help the user explore the portions of an information space relevant to her/his information needs?}

This is particularly useful in Geographic Information Retrieval \cite{Jones-Purves:08,Ballatore-etal:16}, where several queries are aimed at finding the available items, per category, within a geographical area.
Guiding the user towards the exploration of information, such as in the above examples, enables her or him to quickly generate custom maps reflecting individual information needs.

\begin{table*}
\centering
\begin{tabular}{|l|l|l|l|l|}
\hline
\textbf{AnonID} & \textbf{Query}                      & \textbf{QueryTime}  & \textbf{ItemRank} & \textbf{ClickURL}    \\ \hline
67910  & las vegas sports teams                    & 2006-04-30 18:14:57 & 1.0      & http://www.vegas.com                    \\ \hline
67910  & las vegas transportation                  & 2006-04-30 18:19:59 & 1.0      & http://www.vegas.com                    \\ \hline
67910  & mccarran international airport            & 2006-04-30 18:22:30 & 3.0      & http://en.wikipedia.org                 \\ \hline
67910  & mccarran international airport            & 2006-04-30 18:22:30 & 1.0      & http://www.mccarran.com                 \\ \hline
67910  & hub airports in the united states         & 2006-04-30 18:25:28 & 9.0      & http://www.airportcodes.us              \\ \hline
67910  & hub airports in the united states         & 2006-04-30 18:25:28 & 9.0      & http://www.airportcodes.us              \\ \hline
67910  & black las vegas itineraries               & 2006-04-30 18:30:03 &          &                                         \\ \hline
67910  & educational facilities in las vegas       & 2006-04-30 18:30:53 &          &                                         \\ \hline
67910  & medical facilities in las vegas nv        & 2006-04-30 18:31:44 & 1.0      & http://lasvegas.citysearch.com          \\ \hline
67910  & medical facilities in las vegas nv        & 2006-04-30 18:31:44 & 3.0      & http://www.lasvegasnevada.gov           \\ \hline
67910  & medical facilities in las vegas nv        & 2006-04-30 18:31:44 & 9.0      & http://www.lasvegasrelocating.com       \\ \hline
67910  & unique architecture in las vegas nv       & 2006-04-30 18:35:02 & 10.0     & http://www.guggenheim.org               \\ \hline
67910  & unique architecture in las vegas nv       & 2006-04-30 18:35:02 & 2.0      & http://travel.yahoo.com                 \\ \hline
67910  & architecture in las vegas nv              & 2006-04-30 18:40:28 & 4.0      & http://lasvegas.citysearch.com          \\ \hline
67910  & architecture in las vegas nv              & 2006-04-30 18:40:28 & 2.0      & http://www.library.unlv.edu             \\ \hline
67910  & architecture in las vegas nv              & 2006-04-30 18:40:28 & 3.0      & http://local.yahoo.com                  \\ \hline
67910  & religious sites in lasvegas              & 2006-04-30 18:50:35 & 1.0      & http://www.lasvegas.worldweb.com        \\ \hline
67910  & religious sites in lasvegas              & 2006-04-30 18:50:35 & 2.0      & http://www.lasvegas.worldweb.com        \\ \hline
67910  & religious sites in lasvegas              & 2006-04-30 18:56:09 & 12.0     & http://travel2.nytimes.com              \\ \hline
67910  & religious sites in lasvegas              & 2006-04-30 18:56:09 & 12.0     & http://travel2.nytimes.com              \\ \hline
\end{tabular}
\caption{Sample session from the AOL log.}
\label{t:AOLlog}
\end{table*}

In order to answer our research question, we developed a session-based concept suggestion model which, starting from the queries submitted by the user, proposes a set of possibly relevant data categories, complementing the types of information that he or she has focused on. 
Our model is based on an ontological representation of the information space, and on a Natural Language approach to the interpretation of search queries that identifies the referred concepts in a flexible way, by considering synonyms and by applying Word Sense Disambiguation to resolve the meaning of words.
For the development of our model, we analyzed the log of a major search engine and we generated a weighted co-occurrence graph that represents how often concepts are searched together in the sessions of the log. Then, we extracted the clusters of concepts that most frequently co-occur by applying a community detection algorithm to the graph. Those clusters are the basis for query expansion, starting from the concepts referred by the user in the observed part of the search session.
We defined a few heuristics for recommending relevant types of information and we tested them on the log, achieving satisfactory accuracy results.

The main contributions of our work are: (i) a session-based query expansion model that suggests complementary concepts for satisfying the user's information needs in geographic exploratory search, and (ii) evaluation results of the model.   

The following section positions our work in the related research. Section DATASET describes the dataset we used for our experiments. Section IDENTIFYING CLUSTERS OF FREQUENTLY OCCURRING CONCEPTS describes the approach we adopted to identify the co-occurrences of concepts in search sessions. Section SESSION-BASED QUERY EXPANSION proposes some query expansion strategies and the subsequent section evaluates the strategies. The last section concludes the paper and outlines some future work.

\section{Related Work}
\label{related}


Various semantic information retrieval models employ concept networks to identify the meaning of queries and propose query expansions aimed at finding information, regardless of the terminology used by the user. E.g., both \cite{Qiu-Frei:93} and \cite{Grootjem-vanDerWeide:06} use a local thesaurus inferred from the source pool of documents to identify the concepts referred by the queries. Moreover, \cite{Mandala-etal:99} shows that the integration of different types of thesauri (linguistic, domain specific, etc.) improves the performance of query expansion techniques. The co-occurrence of word phrases in documents is also mined in \cite{Berg-Schuemie:99} and \cite{Hoeber-etal:05} to automatically generate associative conceptual spaces, and in \cite{Joshi-Motwani:06} and \cite{Akaishi-etal:04} for term suggestion based on the documents returned by search engines. Differently, Wang et al. \cite{Wang-etal:12} classify queries in patterns according to their syntactic components and match them to a knowledge base to generate the answers. Finally, a knowledge-based approach is adopted in \cite{Wang-etal:17} for concept interpretation, and in \cite{Fernandez-etal:11} to enhance information retrieval in the Semantic Web. On a different perspective, \cite{Molina-Bayarri:11} proposes domain-specific ontologies for the interpretation of queries, assuming that users find it easier to specify what they want to do, rather than the concepts they are interested in. 

Similarly, we use an ontology, and linguistic information, to interpret search queries at the conceptual level. However, we offer a {\em ``you can also be interested in"} function to propose complementary concepts, i.e., topics, for expanding queries in order to satisfy the user's information needs, in a serendipitous way. For this purpose, we propose an associative information retrieval model \cite{Giuliano-Jones:62} based on the observation of concepts co-occurrence in search sessions. Our work is related to the contextual query suggestion model by Cao et al. \cite{Cao-etal:08}, who suggest queries on the basis of the context provided by the user's recent search history. However, we mine {\em ontology concepts} from a linguistic interpretation of search queries, while the concepts defined in \cite{Cao-etal:08} are clusters of queries associated to similar sets of click results selected by users. 

Our model also relates to session-based term suggestion approaches such as the one by Huang et al. \cite{Huang-etal:03}, and with term suggestion models used for web site advertisement, e.g., see Chen et al. \cite{Chen-etal:08}. However, it differs from those works because we look for {\em concept co-occurrence}, which abstracts from the specific words used to refer to concepts, while they observe {\em term co-occurrence} for query expansion. 

Some recent work on information filtering and in recommender systems attempts to acquire relations among information items from the observation of users' behavior, and is complementary to our work. E.g., Google search engine manages the Knowledge Graph \cite{GoogleKnowledgeGraph} to relate facts, concepts and entities depending on their co-occurrence in queries. Moreover, Oramas et al. \cite{Oramas-etal:16} use a knowledge graph for personalized item recommendation in the music domain. Furthermore, CoSeNa \cite{Candan-etal:09} employs keyword co-occurrence in the corpus of documents to be retrieved, and ontological knowledge about the domain concepts, to support the exploration of text collections using a keywords-by-concepts graph. 
Our work differs from the above mentioned ones because we exploit a knowledge graph to predict further concepts that the user might be interested in, i.e., we suggest topics, not individual items.

Working at the conceptual level, our work is also complementary to the research on exploration vs. exploitation in information retrieval, which provides models to recognize the type of search that the user is performing, and/or to adapt the search results accordingly; e.g., \cite{Porrini-etal:14,Athukorala-etal:16,Medlar-etal:17}.

\section{Dataset}
\label{data}
We defined our concept suggestion model, and evaluated its accuracy, by using as a dataset the AOL query log.\footnote{We retrieved the AOL query log in June 2016 from http://www.cim.mcgill.ca/$\tilde{}$dudek/206/Logs/AOL-user-ct-collection/. \newline Currently, the log is available at\newline https://archive.org/details/AOL\_search\_data\_leak\_2006.} 
The log is composed of 10 files; however, we excluded one of them because it seemed to be corrupted by queries likely performed by a bot: that file included a session spawning over 6 days and included more than 20000 queries, or click through, submitted by the same user.

Each line of the AOL log represents either a query, or a click-through event (in which the user clicks on one of the search results of the previously submitted query). The lines contain the following fields:
\begin{itemize}
\item \textit{AnonID}: ID that represents a user in an anonymous way.
\item \textit{Query}: search query submitted by the user.
\item \textit{QueryTime}: date and hour when the query was submitted.
\item \textit{ItemRank}: given the list of results (links to web pages) returned by the search engine, this field represents the relative position of the link on which the user has clicked. If the user has not selected any search results, this field is empty. 
\item \textit{ClickURL}: URL of the search result selected by the user. Similar to field \textit{ItemRank}, this field can be empty.
\end{itemize}
Table \ref{t:AOLlog} shows a session from the AOL log. It can be noticed that the user looks for various types of information, concerning sports, transportation, medical facilities, and others. This substantiates the need for a suggestion model that helps her/him by proposing different topics for exploration.

Since we aim at developing an intelligent search support function for Geographical Information Systems, we selected an example system for our experiments, i.e., OnToMap \cite{Ardissono-etal:17d,Ardissono-etal:17b}, 
and we pre-processed the AOL log to work on a smaller dataset, focused on the search sessions relevant to geographical information. The ultimate goal is to train a query expansion model suitable for improving information search in OnToMap.

In order to build the dataset of our experiments, we first identified the search sessions from the log: we aggregated the queries performed by the same user according to their temporal proximity, following the widely applied rule that two consecutive queries belong to different sessions if the time interval between them exceeds half an hour; see \cite{White-etal:07}.
Then, we selected the sessions including at least one query that refers to the concepts of the OnToMap ontology.\footnote{We did not consider any broader ontologies, such as WordNet \cite{WordNet} or ProBase \cite{Wu-etal:12}, because we aimed at obtaining a focused dataset to be analyzed.} 
For this purpose, our main task was the identification of the concepts referred by the terms of the queries, which we carried out using the approach described in \cite{Ardissono-etal:16,Mauro-Ardissono:17b}. Concept identification is a particularly important task: as reported in \cite{Wang-etal:12}, the analysis of the log of a major search engine proved that about the 62\% of the queries contain at least one conceptual class term. 

\begin{table}
\centering
\begin{tabular}{|l|l|}
\hline
\textbf{rdfs:label} & "Kindergarten"@en     \\ \hline
\textbf{rdfs:comment} & \makecell[l]{"An educational institution for young \\children, usually before they go to \\primary school."@en}  \\ \hline
\textbf{:keywords} & \makecell[l]{"child"@en,  "educational"@en,\\ "young"@en}  \\ \hline
\textbf{:lemma} & "kindergarten"@en \\ \hline
\textbf{:synonyms} & \makecell[l]{"0th\_grade"@en,  \\"all-day\_kindergarten"@en, \\"didactics"@en, "childcare"@en, \\"educability"@en, \\"education\_program"@en,\\ "education\_system"@en, \\"junior\_kindergarten"@en, \\"kindergarden"@en, \\"nursery"@en, "pre-primary"@en, \\"pre-school"@en, "preschoolar"@en}  \\ \hline
\end{tabular}
\caption{Ontology definition of concept "Kintergarden". Values are tagged with the reference language (@en, for English).}
\label{t:kintergarden}
\end{table}

\subsection{Reference Ontology}

The OnToMap ontology describes concepts by merging linguistic, encyclopedic and technical knowledge, focusing on concepts that are typical of Participatory Processes; see \cite{Ardissono-etal:16,Voghera-etal:16}. 
Each concept has a textual description, and a set of lemmatized synonyms and keywords extracted from the description. E.g., Table \ref{t:kintergarden} shows the specification of concept "Kintergarden". It reports the concept name (rdfs:label), its textual description (rdfs:comment), a few lemmatized keywords (:keywords), the lemma of the concept (:lemma), and a set of lemmatized synonyms of the terms occurring in the description (:synonyms).

\subsection{Creation of the Dataset for the Experiments}

We identified the AOL queries relevant to our experiments by matching the knowledge about concepts (i.e., the lemmatized synonyms and keywords extracted from the textual description of the ontology concepts) to the lemmatized words that compose the queries. If there was at least one match between a concept and a query, we considered the query as relevant, and we included the session to which it belonged in our dataset.

Each query can refer to one or more ontology concepts: the mean number of concepts per query of AOL-reduced is 1.502. This is due to the following reasons: 
\begin{itemize}
\item  
The query might refer to independent concepts, each one identified in an unambiguous way. For instance, in a sample query like "public school and transportation in New York" the identified concepts would be \textit{School} and \textit{Local Public Transportation}. Notice that, similar to the findings reported in \cite{Beitzel-etal:05}, most queries of the AOL log are short and refer to a single concept, as in the examples of Table \ref{t:AOLlog}. 
\item
Because of ambiguity issues, more than one concept could be identified. For instance, given query "missouri child support" from the AOL log, the concepts identified from word "child" are \textit{Childcare Service}, \textit{Play Area} and \textit{Kindergarten}, both including term "child" in their descriptions. As the query is short, it is difficult to understand which topic is the most probable one. Therefore, we assume that the query matches all concepts, with uncertainty. 
\newline
In the analysis of the log, we could not exploit the search results selected by users to disambiguate the queries, because we found out that, in most cases, they refer to the root pages of large web sites (e.g., see the URLs in Table \ref{t:AOLlog}), or they are obsolete (e.g., the link reported as a search result of query "missouri child support", http://www.dss.state.mo.us). Thus, they help in few cases. 
\end{itemize}

\begin{figure}
  \centering
  \includegraphics[width=0.9\linewidth]{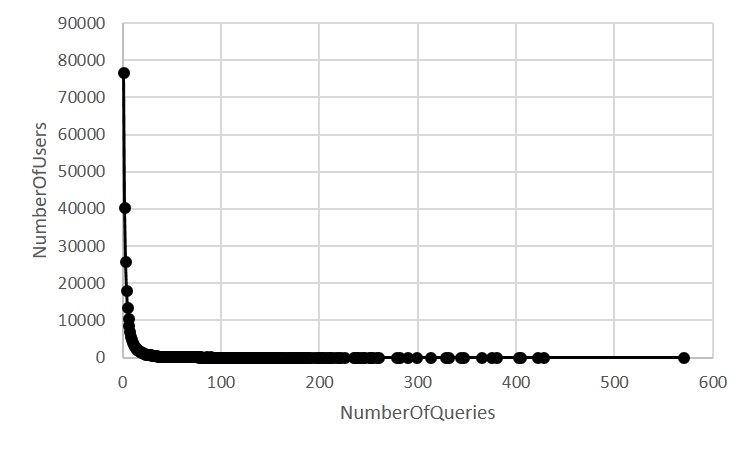}
  \caption{Distribution of the number of queries per user.}\label{fig:distribuzione-query-utenti}
\end{figure}

\begin{table}[]
\centering
\begin{tabular}{|l|l|}
\hline
\textbf{Minimum number of queries per user} & 1     \\ \hline
\textbf{Maximum number of queries per user} & 428   \\ \hline
\textbf{Mean number of queries per user} & 6.38  \\ \hline
\textbf{Median number of queries per user}                             & 3     \\ \hline
\textbf{Standard Deviation}                 & 11.37 \\ \hline
\end{tabular}
\caption{Measures about the distribution of the queries per user.}
\label{t:queries-users}
\end{table}

\begin{figure}
  \centering
  \includegraphics[width=0.9\linewidth]{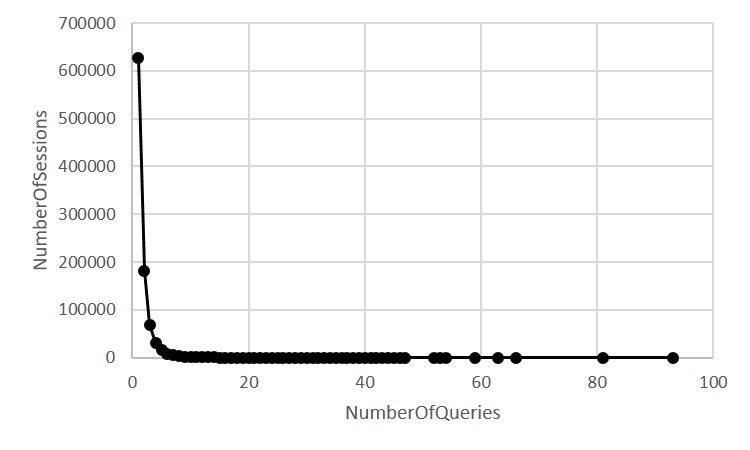}
  \caption{Distribution of queries in  sessions.}\label{fig:distribuzione-lunghezza-sessioni}
\end{figure}

\begin{figure}[b]
  \centering
  \includegraphics[width=0.9\linewidth]{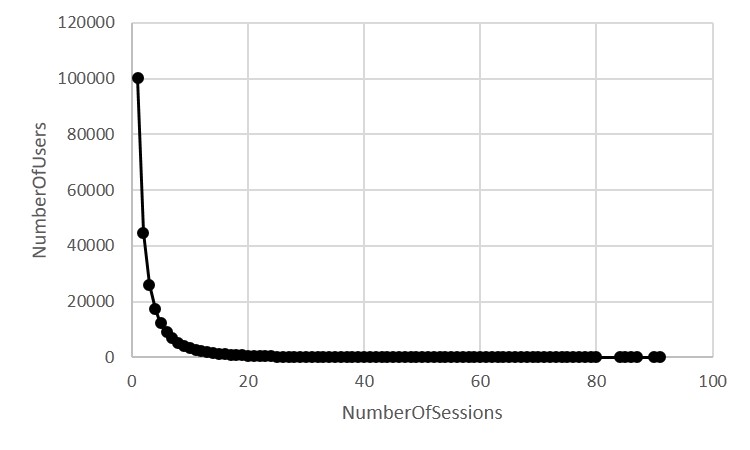}
  \caption{Distribution of sessions per users.}\label{fig:distribuzione-sessioni-utenti}
\end{figure}

\begin{table}[]
\centering
\begin{tabular}{|l|l|}
\hline
\textbf{Minimum number of queries per session} & 1    \\ \hline
\textbf{Maximum number of queries per session} & 93   \\ \hline
\textbf{Mean number of queries per session} & 1.67 \\ \hline
\textbf{Median number of queries per session} & 1    \\ \hline
\textbf{Standard Deviation}                    & 1.45 \\ \hline
\end{tabular}
\caption{Measures about the length of the search sessions.}
\label{t:lunghezza-sessioni}
\end{table}
\subsection{Characteristics of the Dataset}
\label{dataset}

The dataset obtained after the pre-processing phase, henceforth denoted as {\em AOL-reduced}, is composed of 1581817 queries submitted by 247868 users.
The chart in Figure \ref{fig:distribuzione-query-utenti} shows the distribution of users with respect to the number of queries they specified. It can be noticed that the distribution follows a Power Law: most users submitted a very small set of queries, whereas a small group of users submitted several queries. As shown in Table \ref{t:queries-users}, the mean number of queries per user is 6, but the distribution differs from this value for about 11 queries (standard deviation).

The dataset includes 945945 sessions, on which we performed two types of analyses:
\begin{itemize}
\item
Firstly, we analyzed the distribution of sessions w.r.t. the number of queries they include.
Table \ref{t:lunghezza-sessioni} shows the measures related to the length of sessions considering the number of queries. The mean length of a session is 1.67.
The distribution follows a Power Law, as shown in Figure \ref{fig:distribuzione-lunghezza-sessioni}.
\item
Secondly, we analyzed the distribution of sessions with respect to users in order to compute the mean number of sessions that the users started; see Figure \ref{fig:distribuzione-sessioni-utenti}. We found out that users engaged in a mean number of about 3.82 sessions (median = 2). Moreover, the distribution follows a Power Law, with several people engaging in few search sessions, and a small number of users starting a larger number of sessions.
\end{itemize}

\section{Identifying clusters of frequently co-occurring concepts}
\label{clusters}

As described in the introduction, starting from the interpretation of the search queries that the user submits, we aim at suggesting further concepts which could be relevant to her/his needs. The idea is to provide an adaptive list of pointers to the types of information provided by the system, which are useful in the context of the user's search. 
We did this by identifying, from the concepts referred by the queries of the AOL-reduced dataset, other concepts that are related to the former from the viewpoint of commonly shared interests.


\subsection{Step 1: Creation of the Concept Co-occurrence Graph}
\label{graph}
Starting from the AOL-reduced dataset, we built a graph that represents concepts co-occurrence in search sessions. 
By co-occurrence we mean the fact that two or more concepts are referred by the queries belonging to the same session.
The graph is composed as follows:
\begin{itemize}
\item Each node represents a concept;
\item Each edge represents the co-occurrence weight of the two connected concepts. This weight is computed by summing up the evidence of co-occurrence observed in all the sessions of the dataset.
\end{itemize}
The idea is that, each time two concepts are identified within the same session, the weight of the edge connecting them must increase, in order to capture and reinforce the hypothesis that people frequently search these concepts together.

Given two concepts $c_i$ and $c_j$, the weight of the edge that connects them is defined as:
\begin{equation}
\label{eq1}
w_{{c_i}{c_j}}=\sum_{S=1}^{nSessions} Freq_{S_{ij}}
\end{equation}
where $Freq_{S_{ij}}$ represents the contribution provided by session $S$ to the co-occurrence frequency of $c_i$ and $c_j$. 

The contribution of the sessions to the weights of the co-occurrence graph ($CG$) is represented by a local weighted graph created by interpreting the queries of $S$.
Specifically, $Freq_{S_{ij}}$ is obtained by considering the evidence of co-occurrence provided by the queries that compose $S$. However, within a session, we avoid summing up the evidence provided by multiple occurrences of the same concepts, because they could derive from a query reformulation \cite{Rieh-Xie:06} or from the repetition of the queries in click-through events; see Table \ref{t:AOLlog}.

Given $S=\{Q_1, \dots, Q_n\}$, 
$Freq_{S_{ij}}$ is thus computed as follows: 
\begin{equation}
\label{eq2}
Freq_{S_{ij}} = Max_{k=1}^{n}(Freq_{ij_{Q_k}}, ev_{ij_{Q_{k-1}}})
\end{equation}
where $Freq_{ij_{Q_k}}$ is the co-occurrence evidence of concepts $c_i$ and $c_j$  provided by query $Q_k$, and $ev_{ij_{Q_{k-1}}}$ is the one estimated during the interpretation of queries $Q_1, \dots, Q_{k-1}$.

Finally, the contribution of a query $Q$ is computed as follows:
\begin{itemize}
\item 
If $Q$ contains $n$ terms ($n>=0$), each one identifying a non-ambiguous concept: 
$T_1 \Rightarrow c_1, \quad  T_2 \Rightarrow c_2, \quad \dots, \quad T_n \Rightarrow c_n$, then, $Q$ generates nodes $c_1, \dots, c_n$ of the local graph (if they do not exist yet). For all pairs of concepts $c_a$ and $c_b$ referred by $Q$, $Freq_{ab_{Q}} = 1$. Moreover,
$Freq_{vw_{Q}} = 1$ for all the edges connecting $c_a$, or $c_b$, to some non-ambiguous concept of the local graph. $Freq_{vw_{Q}} = 0$ for the other edges of the graph.

For instance, suppose that $Q = $"public school and transportation in New York" is the first query of a session. The terms and concepts referred by $Q$ are:
\begin{itemize}
\item
school $\Rightarrow \textit{School}$ - denoted as concept $x$;
\item
transportation $\Rightarrow \textit{Local Public Transportation}$ - denoted as concept $y$.
\end{itemize}
Then, the local graph is composed of the $x$ and $y$ nodes. Moreover, $Freq_{xy_{Q}} = 1$. 
\item 
If $Q$ contains a term $t$ that refers to $m$ concepts $\{c_1, \dots,c_m\}$, the interpretation is ambiguous. Therefore, the evidence brought by $t$ to the concepts is $\frac{1}{m}$, in order to take into account the possible interpretations of the query, and spread the weight to the ambiguous concepts. Thus, for each pair $c_a$, $c_b$ in $\{c_1, \dots,c_m\}$, $Freq_{ab_{Q}} = \frac{1}{m}$.
$Freq_{vw_{Q}} = 1/m$ for all the edges connecting $c_a$, or $c_b$, to the other concepts of the local graph. 
Finally, $Freq_{vw_{Q}} = 0$ for all the edges of the local graph that are not outgoing arcs of $c_a$ or $c_b$.  

For example, if $Q = $ "missouri child support" is the first query of a session: 
\begin{itemize}
\item  
The concepts identified from word "child" are: \textit{Childcare Services}  - denoted as $x$, \textit{Play Areas} - $y$, and \textit{Kindergartens} - $z$.
\item
Then, the local graph is composed of the $x$, $y$ and $z$ nodes. Moreover,
$Freq_{xy_{Q}} = Freq_{xz_{Q}} = Freq_{yz_{Q}} = \frac{1}{3}$.
\end{itemize}
\end{itemize}
We now sketch the generation of the local co-occurrence graph for a sample search session $S$. We recall that the contribution of multiple sessions to the overall co-occurrence graph ($GC$) is incremental: i.e., we sum up the weights provided by the individual sessions.
Let's suppose that the terms of $S$ refer to the following concepts:
\newline
$Q_1: t1 \Rightarrow c_1$ \newline
$Q_2: t2 \Rightarrow c_2$ \newline
$Q_3: t3 \Rightarrow c_3, c_4$ \newline
$Q_4: t4 \Rightarrow c_2$ \newline
$Q_5:t5 \Rightarrow c_3$

\begin{figure}
  \centering
  \includegraphics[width=0.5\linewidth]{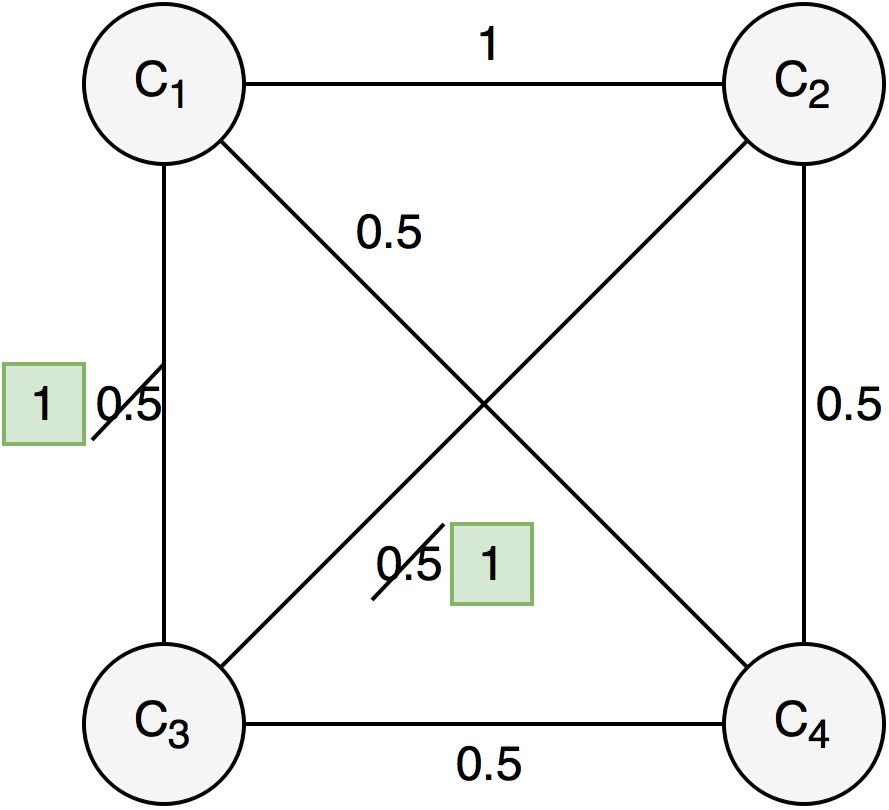}
  \caption{Construction of the graph describing the concepts co-occurrence frequency for a session.}\label{fig:esempio-grafo}
\end{figure}

Figure \ref{fig:esempio-grafo} shows the local co-occurrence graph generated by $S$.
\begin{itemize}
\item 
$Q_1$ adds node $c_1$ to the local graph.
\item
$Q_2$ adds $c_2$ and assigns a weight = 1 to the edge between $c_1$ and $c_2$, because there is no ambiguity: 
\newline
$Max(Freq_{{c_{1}c_{2}}_{Q_2}}, ev_{{c_{1}c_{2}}_{Q_1}}) = Max(1, 0) = 1$.
\item  
$Q_3$ adds $c_3$ and $c_4$. Moreover, it assigns a weight = $\frac{1}{2}$ to the edge connecting $c_3$ and $c_4$ (the ambiguity concerns two concepts). Furthermore, it assigns the same weight to the edges connecting $c_3$, and $c_4$, to the concepts of the previous queries (i.e., the edges between $c_3$ and $c_1$, $c_3$ and $c_2$, $c_4$ and $c_1$, $c_4$ and $c_2$).
\item
$Q_4$ does not add any nodes, nor does it modify the weights in the local graph because the evidence of $c_2$ doesn't solve the ambiguity between $c_3$ and $c_4$:\newline
$Max(Freq_{{c_{2}c_{3}}_{Q_4}}, ev_{{c_{2}c_{3}}_{Q_3}}) = Max(0.5, 0.5) = 0.5$ \newline
$Max(Freq_{{c_{2}c_{4}}_{Q_4}}, ev_{{c_{2}c_{4}}_{Q_3}}) = Max(0.5, 0.5) = 0.5$ \newline
$Max(Freq_{{c_{1}c_{2}}_{Q_4}}, ev_{{c_{1}c_{2}}_{Q_3}}) = Max(1, 1) = 1$.
\item  
$Q_5$ does not add any nodes, but it solves the ambiguity between $c_3$ and $c_1$, and $c_3$ and $c_2$, respectively, because it provides a non ambiguous evidence of $c_3$. Thus, the respective weights of the local graph are updated:
\newline 
$Max(Freq_{{c_{1}c_{3}}_{Q_5}}, ev_{{c_{1}c_{3}}_{Q_4}}) = Max(1, 0.5) = 1$.
\newline
$Max(Freq_{{c_{2}c_{3}}_{Q_5}}, ev_{{c_{2}c_{3}}_{Q_4}}) = Max(1, 0.5) = 1$.
\newline
Notice that we maintain the ambiguity between $c_3$ and $c_4$ (in a conservative approach) because, in general, we cannot assume that $Q_5$ is a reformulation of $Q_3$. Thus, we do not use it to exclude $c_4$ from the interpretation of $Q_3$.
\end{itemize}

\begin{figure}[b]
  \centering
  \includegraphics[width=0.9\linewidth]{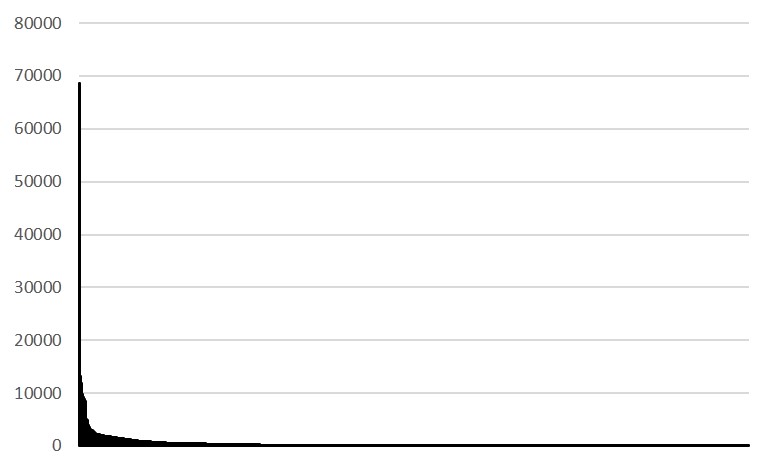}
  \caption{Distribution of the weight of edges in the co-occurrence graph.}\label{fig:distribuzione-grafo}
\end{figure}

\subsection{Step 2: Pruning the Graph}
\label{pruning}

The co-occurrence graph of the AOL-reduced dataset is strongly connected: almost all of the ontology concepts are linked to each other by an edge whose weight is $>0$. 
We thus decided to analyze the distribution of weights in the graph, in order to understand the strength of the correlation between concepts.
Figure \ref{fig:distribuzione-grafo} shows this distribution: the x-axis represents the edges, and the y-axis represents their weights, which take values in [0, 68707.5]. 

The distribution highlights the fact that there is a large number of weakly connected concepts, i.e., of candidate items for query expansion. 
As, different from keyword suggestion in search engines, we aim at proposing few concepts, we decided to delete the edges having low weight, assuming that they represent weak associations between concepts, and that they capture less commonly shared interests. 

We selected a threshold for pruning the graph in order to optimize the prediction accuracy of the resulting concept co-occurrence clusters; i.e., the degree of matching between the set of concepts composing the clusters and the concepts identified in the sessions of AOL-reduced. The selected threshold is 2200. Details about the evaluation of accuracy follow.

\subsection{Step 3: Creation of the Concept Co-occurrence Clusters}
\label{communities}

Starting from a co-occurrence graph ($CG$) pruned with a threshold, the clusters are created by applying a community detection algorithm which identifies the sets of strictly correlated concepts. Our hypothesis is that they correspond to sets of concepts that are frequently searched together.  

For the identification of clusters we analyzed various algorithms. We selected COPRA \cite{Gregory:10}, which works on weighted graphs and detects overlapping communities. In this way, a concept can belong to different clusters at the same time. Our idea was that, within a session, the user might focus on more than one set of highly correlated concepts, and that, starting from the same query, (s)he might explore different paths of the information space. 

The COPRA algorithm, applied to the concept co-occurrence graph pruned with threshold 2200, returned 23 clusters having a minimum and maximum number of concepts equal to 1 and 6, respectively. 

Three sample clusters are:
\begin{enumerate}
\item \{Play Area, Sport Area, Kindergarten, Law Enforcement, Hospital\}.
\item \{National Park, Provincial Park, Regional Park, Urban Park, Furnished Green\}.
\item \{Play Area, Sport Area, Library, Childcare service, Law Enforcement, School\}.
\end{enumerate}

\subsection{Step 4: Validation of the Clusters}
\label{cluster validation}

We tested the clusters returned by COPRA on different versions of the pruned graph (using different weight thresholds) in order to estimate how close they reflected user behavior in the AOL-reduced dataset. We considered the AOL-reduced sessions as the ground-truth and we computed standard accuracy measures for the evaluation. 

We treated the clusters as unordered sets and we did not care about the observed order of exploration of concepts in the sessions. The reason for this is the fact that, after each search query, we plan to suggest concepts as a set of selectable items, in a multi-choice box. This set is a projection on the information categories managed by the system, based on the search context. 

For each threshold, we tested the accuracy of the clusters by applying 10-fold cross-validation, after having randomly distributed the sessions of the dataset on folders. We used 90\% of the sessions as learning set and 10\% as test set. 

Given a session $S = \{c_{s_1}, \dots, c_{s_n}\}$, and a cluster $CL = \{c_{cl_1}, \dots, c_{cl_m}\}$, we evaluated the following measures:
\begin{itemize}
\item
Precision, describing the rate of concepts of $CL$ that also occur in $S$. These concepts represent correct predictions if $CL$ is used for query expansion: 
\begin{equation}
precision = \frac{|\{c_{s_1}, \dots, c_{s_n}\} \cap \{c_{cl_1}, \dots, c_{cl_m}\}|}{|\{c_{cl_1}, \dots, c_{cl_m}\}|}
\end{equation}
\item
Recall, representing the rate of relevant concepts of $S$ which $CL$ contains, and thus can suggest, if used for query expansion:
\newline 
\begin{equation}
recall = \frac{|\{c_{s_1}, \dots, c_{s_n}\} \cap \{c_{cl_1}, \dots, c_{cl_m}\}|}{|\{c_{s_1}, \dots, c_{s_n}\}|}
\end{equation}
\item
F1 score, computed as: 
\begin{equation}
2 * \frac{precision * recall}{precision + recall} 
\end{equation}
\end{itemize}
Moreover, we considered two types of evaluation:

\begin{table}[b]
\centering
\begin{tabular}{|l|l|l|}
\hline
                   & \textbf{Eval1} & \textbf{Eval2} \\ \hline
\textbf{Precision} & 0.659       & 0.626       \\ \hline
\textbf{Recall}    & 0.794      & 0.794       \\ \hline
\textbf{F1}        & 0.720       & 0.700      \\ \hline
\end{tabular}
\caption{Evaluation of the clusters accuracy.}
\label{t:clusters-accuracy}
\end{table}

\begin{itemize}
\item Eval1: accuracy of the clusters that best represent search behavior within a session. The aim of this test was to measure the optimal adherence of the clusters to users behavior, assuming to be able to identify the best cluster from the observed queries. In this case, for each session $S$, we computed the precision, recall and F1 score of all the clusters. Then, we selected the accuracy of the best performing one as the representative value for $S$. Finally, we computed the mean accuracy of such best clusters.

\item Eval2: mean accuracy of the clusters that include at least one concept referred in the search session.
In this case, we aimed at testing the accuracy of a broader set of clusters, related to search sessions in a looser way. As above, we computed the mean precision, recall and F1 score.
\end{itemize}
Table \ref{t:clusters-accuracy} reports the accuracy values we obtained using the threshold that maximizes the mean F1 ($threshold = 2200$) for pruning the concept co-occurrence graph: 
\begin{itemize}
\item  
The precision values show that, during the search sessions, users refer to a substantial portion of the concepts included in the clusters. However, the clusters contain some concepts that are not explored: by using all the clusters that have at least one concept in common with the session, we obtain a precision of about 0.626. 
\item  
The recall shows that the suggested clusters largely cover the search sessions: in both evaluations, the recall is about 0.794. Thus, the clusters help propose relevant concepts. 
\end{itemize}
Overall, we can say that, by selecting clusters that have some concepts in common with those referred in the observed portion of a search session, we have good chances to suggest types of information that the user will be interested in.
These results helped us in the identification of the query expansion strategies described in the following section.

In the computation of these measures, we did not take into account the occurrence of ambiguities in the interpretation of search queries. Basically, we considered all the concepts referred (either ambiguously or non ambiguously) in the queries as concepts belonging to the search sessions. This might introduce some noise in the evaluation results (e.g., it could increment the number of concepts relevant to the search sessions, reducing the precision results). However, we could not retrieve more precise information about the search interests of the AOL users because, as already discussed, most queries of the dataset are short, and the dataset provides little information to disambiguate them.

\section{Session-based query expansion}
\label{recommendation}

Let's consider a search session $S = \{Q_1, \dots, Q_i, \dots, Q_n\}$, and suppose that we have observed the first $i$ queries of $S$. 
We denote the sequence $\{Q_1, \dots, Q_i\}$ as $S@i$, and the set of concepts identified by interpreting the queries of $S@i$ as:
\newline 
$C@i = \{c1, \dots, \{c_{k_1}, \dots, c_{k_m}\}, \dots, \{c_{t_1}, \dots, c_{t_s}\}, \dots, c_n\}$. 
\newline 
In $C@i$, the ambiguities in concept identification are represented by including the tuples of ambiguous concepts in subsets of $C@i$. For instance, in the above example, $\{c_{k_1}, \dots, c_{k_m}\}$ and, respectively, $\{c_{t_1}, \dots, c_{t_s}\}$ are ambiguous. In the following, we will refer to the tuples of ambiguous concepts as {\em ambiguity sets}.

We consider three concepts suggestion strategies, which differ from each other in the method for selecting the clusters to be used for query expansion.
We assume that they could be applied immediately after the interpretation of the first query of a session, or incrementally, in order to support interactive information search.

\begin{itemize}
\item
{\bf SLACK:} 
\begin{enumerate}
\item
Select a set of clusters $\{CL_{x_1}, ..., CL_{x_y}\}$ that contain {\em at least} one concept $c$ such that $c \in C@i$; i.e., $c$ has been referenced in the observed portion of the search session, $S@i$, either ambiguously or unambiguously. 
\item
Propose the concepts that belong to at least one of the selected clusters, and the user has not yet explored. For each cluster $CL_{x_j}$, we denote the set of concepts included in $CL_{x_j}$ as $Sugg_{x_j}$. The set of concepts to be suggested is computed as follows: 
\newline
$Sugg@i = \{Sugg_{x_1} \cup \dots \cup Sugg_{x_y}\} - C@i$. 
\end{enumerate}
For instance, given $C@i = \{c_1, \{c_3, c_4\}\}$, and two selected clusters, $CL_1 = \{c_1, c_7\}$ and $CL_2 = \{c_2, c_3, c_5, c_8\}$, $Sugg@i =\{c_2, c_5, c_7, c_8\}$.
\item
{\bf SLACK-selective:}
Same as SLACK, but only the clusters that best match $C@i$ are used to compute $Sugg@i$. 
For our experiments, we selected the best matching cluster. The evaluation of the concept suggestion strategies, described later on, provided satisfactory accuracy results for this setting of the strategy.

We compute the degree of matching between a cluster $CL$ and $C@i$ as the cardinality of $CL \cap C@i$ (i.e., the number of concepts they have in common), taking the ambiguity in $C@i$ into account. Specifically:
\begin{itemize}
\item
Each concept $c$ occurring both in $CL$ and, as a non-ambiguous element, in $C@i$, contributes to the computation of the degree of matching with a value = 1.
\item
Each concept $c$ that occurs in $CL$, and is part of an ambiguity set $AMB$ in $C@i$, contributes to the computation with $\frac{1}{|AMB|}$; i.e., the cardinality of the ambiguity set mitigates its contribution to the computation of the degree of matching.
\end{itemize}
For instance, given $C@i = \{c_1, \{c_3, c_4\}, c_5\}$, the degree of matching with cluster $CL_1 = \{c_1, c_7\}$ is 1 and the one with $CL_2 = \{c_2, c_3, c_5, c_8\}$ is 1.5.
\item
{\bf STRICT:}
\begin{enumerate}
\item
Select the set of clusters $\{CL_{x_1}, \dots, CL_{x_y}\}$ containing {\em all} of the concepts $c$ such that $c \in C@i$. These are the clusters covering the whole portion of the search session observed so far (regardless of the fact that the concepts are ambiguous or not). 
\item
As above, the set of concepts suggested for query expansion is: 
$Sugg@i = \{Sugg_{x_1} \cup \dots \cup Sugg_{x_y}\} - C@i$. 
\end{enumerate}
We did not propose a STRICT-selective strategy because STRICT generates very small candidate sets. 
\end{itemize}

\begin{table}[]
\centering
\begin{tabular}{|l|l|l|l|}
\hline
\textbf{Indicator} & \textbf{SLACK} & \makecell[l]{\textbf{SLACK-}\\\textbf{selective}} & \textbf{STRICT}   \\ \hline
 \makecell[l]{\textbf{Min number of }\\\textbf{candidate clusters}} & 0 & 0 & 0   \\ \hline
 \makecell[l]{\textbf{Max number of }\\\textbf{candidate clusters}} & {\bf 2} & N=1 & 2   \\ \hline
 \makecell[l]{\textbf{Mean number of }\\\textbf{candidate clusters}} & {\bf 1.192} & 1 & 1.188
 \\ \hline
\makecell[l]{\textbf{Min number of }\\\textbf{suggested concepts}} & 0 & 0 & 0   \\ \hline
\makecell[l]{\textbf{Max number of }\\\textbf{suggested concepts}} & {\bf 7} & 5 & {\bf 7}   \\ \hline
\makecell[l]{\textbf{Mean number of }\\\textbf{suggested concepts}} & {\bf 2.688} & 2.309 & 2.681  \\ \hline
\textbf{Precision} &  0.614 & {\bf 0.621} & 0.615   \\ \hline
\textbf{Recall} & {\bf 0.791} & 0.783 & 0.790  \\ \hline
\textbf{F1} & {\bf 0.692} & {\bf 0.692} & {\bf 0.692}   \\ \hline
\textbf{Success rate (\%)} &  {\bf 51.5} & 51.2 & {\bf 51.5}   \\ \hline
\end{tabular}
\caption{Statistical measures of concept suggestion strategies applied to the first query of the search sessions ($S@1$).}
\label{t:measures1}
\end{table}

\begin{table}[]
\centering
\begin{tabular}{|l|l|l|l|}
\hline
\textbf{Indicator} & \textbf{SLACK} & \makecell[l]{\textbf{SLACK-}\\\textbf{selective}} & \textbf{STRICT}   \\ \hline
 \makecell[l]{\textbf{Min number of }\\\textbf{candidate clusters}} & 0 & 0 & 0   \\ \hline
 \makecell[l]{\textbf{Max number of }\\\textbf{candidate clusters}} & {\bf 2} & N=1 & 2   \\ \hline
 \makecell[l]{\textbf{Mean number of }\\\textbf{candidate clusters}} & {\bf 1.223} & 1 & 1.192  \\ \hline
 \makecell[l]{\textbf{Min number of }\\\textbf{suggested concepts}} & 0 & 0 & 0   \\ \hline
 \makecell[l]{\textbf{Max number of }\\\textbf{suggested concepts}}  & {\bf 7} & 5 & {\bf 7}   \\ \hline
 \makecell[l]{\textbf{Mean number of }\\\textbf{suggested concepts}}  & {\bf 2.861} & 2.446 & 2.799   \\ \hline
\textbf{Precision} &  0.594 & {\bf 0.598} & 0.596
  \\ \hline
\textbf{Recall} & {\bf 0.819} & 0.801 & 0.818   \\ \hline
\textbf{F1} &  {\bf 0.689} & 0.685 & {\bf 0.689} \\ \hline
\textbf{Success rate (\%)} &  {\bf 39.7} & 39.2 & 39.6   \\ \hline
\end{tabular}
\caption{Statistical measures of concept suggestion strategies applied to the first two queries of the search sessions ($S@2$).}
\label{t:measures2}
\end{table}

\section{Evaluation of concept suggestion strategies}
\label{experiments}

We tested the accuracy of our strategies by applying 10-fold cross-validation on the AOL-reduced dataset, after having randomly distributed the sessions on folders. 
For the evaluation, we compared the concepts suggested by the strategies to those explored by users in the search sessions (ground-truth). Specifically, for each search session, we tested the strategies after having interpreted the first query alone ($S@1$), the first two queries ($S@2$), and so forth. In each case, we computed the mean precision, recall and F1 score of the strategies by comparing the concepts of the suggested clusters to the concepts referred in the remainder of the sessions.

\subsection{Evaluation Results}
\label{measures}

Table \ref{t:measures1} summarizes the results that we obtained in the suggestion of concepts immediately after having interpreted the first query of the search sessions ($S@1$):
\begin{itemize}
\item
SLACK: the minimum, mean, and maximum number of candidate clusters identified by this strategy are 0, 1.192, and 2, respectively.  
The mean number of concepts suggested for query expansion is about 2.688. This number goes up to 7 in some sessions.
\item
SLACK-selective with number of best matching clusters N=1. 
The minimum, mean and maximum number of selected clusters are 0, 1 and N=1, respectively. 
The strategy suggests a lower mean number of concepts: about 2.309, and it proposes at most 5 concepts.
\item
The STRICT strategy proposes a maximum of 2 candidate clusters per session, with about 1.188 clusters in average, and it generates at most 7 concept suggestions, with a mean value of about 2.681 concepts. 
\end{itemize}
The three strategies have almost the same accuracy (see Table \ref{t:measures1}): the F1 measure can be approximated to 0.692 in all the cases. However, they differ from each other if we look at finer-grained measures:
\begin{itemize}
\item
Precision: SLACK-selective outperforms the other two, i.e., it suggest the highest number of concepts that users are observed to explore in the remainder of the search session. The lower precision of SLACK can be explained with the fact that it selects a superset of the clusters picked by SLACK-selective, and thus can introduce more noise in the suggestions. Similarly, STRICT can select more than one cluster (covering all the concepts referred in the first query), thus incrementing noise in some cases.
\item
Recall: the best strategy is SLACK, which outperforms SLACK-selective because it selects a larger pool of clusters and thus has better chances to guess the concepts explored by users. Interestingly, however, STRICT achieves almost the same recall, having a stricter cluster selection policy. We explain this finding with the fact that, by selecting all the clusters that cover the observed query, the strategy has good chances to guess the user's exploration path, among the possible ones (remember that the concept co-occurrence clusters mirrored common search behavior in a fairly accurate way).
\item
Success rate: as proposed by Huang et al. in \cite{Huang-etal:03}, this metric measures the percentage of sessions for which our strategies propose at least one concept that is referred in the portion of the search session to be observed. Looking at the table, we can see that all of the strategies are successful in more than 50\% of cases. However, the best performing ones are SLACK and STRICT. In other words, SLACK is as good as STRICT if we aim at suggesting at least one relevant concept.
\newline
We compared our strategies to the Huang et al.'s term suggestion model, which is the most similar one among those available in the literature. As that model achieves 23\% success rate, we can say that our approach definitely outperforms it.  
\end{itemize}
Table \ref{t:measures2} shows the evaluation results after the interpretation of the first two queries of the search sessions.
It can be seen that, for all the strategies, the minimum, mean and maximum number of selected clusters are constant. Moreover, the minimum, mean and maximum number of suggested concepts are constant or increase.
\newline
Regarding accuracy, we notice that for all the three strategies, the recall increases, while the precision and F1 scores decrease to about 0.6, but the SLACK-selective strategy is the most precise one. SLACK has the best recall and STRICT a rather similar one; as result, both strategies have the best F1 value. The higher recall of these strategies suggests that the selection of a single cluster, best representing the observed portion of search session, is not general enough to cover all the possible developments of the search session. Thus, a less restrictive approach, that makes it possible to consider more than one development of a search session (as the one adopted in both SLACK and STRICT) achieves better results.

\begin{figure}
  \centering
  \includegraphics[width=0.9\linewidth]{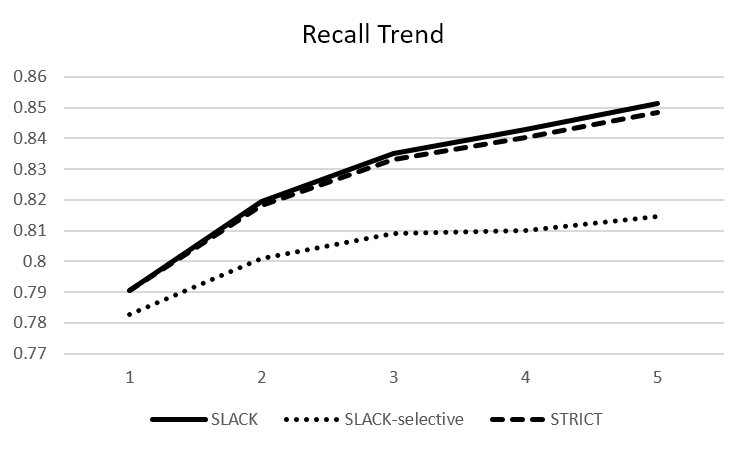}
  \caption{Recall of the concept suggestion strategies in function of the number of interpreted queries ($S@i$)}\label{fig:recall-trend}
\end{figure}

We measured the trend for longer portions of search sessions and we discovered that it is consistent, i.e., the precision decreases, and the recall increases, when more queries are interpreted. See Figures \ref{fig:precision-trend} and \ref{fig:recall-trend}, which report the values of these measures after the interpretation of 1 to 5 queries.\footnote{We did not consider any longer sequences because very few search sessions are composed of more than 5 queries.} However, the three strategies behave slightly differently:

\begin{figure}
  \centering
  \includegraphics[width=0.9\linewidth]{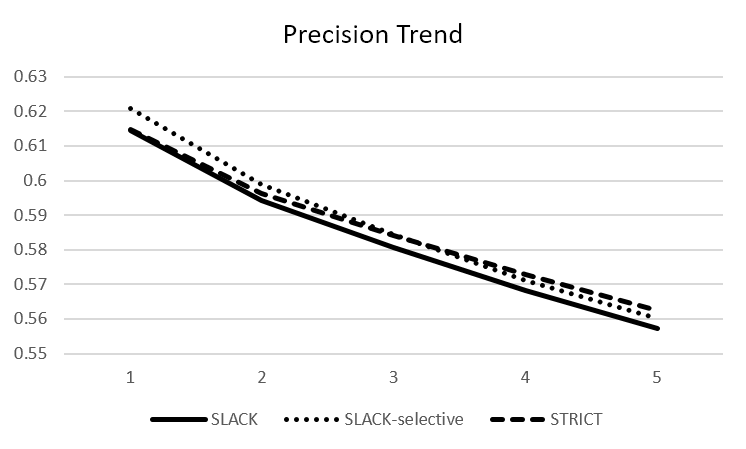}
  \caption{Precision of the concept suggestion strategies in function of the number of interpreted queries ($S@i$).}\label{fig:precision-trend}
\end{figure}

\begin{itemize} 
\item 
SLACK-selective is the most precise strategy when observing the first two queries, and is always better than SLACK, but it is outperformed by STRICT after the observation of 3 queries. We explain this with the fact that, when more information about the search context is available, STRICT is able to select the clusters that best represent the user's search interests, and thus to mirror her/his behavior. 
\item 
Regarding recall, we believe that it increases when more information about a search session is available because, if more concepts are identified from the search queries, there are more chances to select good candidates and thus to propose relevant concepts. However, the improvement of this accuracy measure varies depending on the selectivity of the strategy:
\begin{itemize}
\item 
SLACK selects the highest number of  clusters, and thus suggest a larger set of concepts. This increases the chances that the suggested concepts are identified in the session. 
\item 
SLACK-selective chooses a single cluster that best matches the search session, at the expense of recall.
\item
STRICT selects the clusters matching the whole portion of session observed so far. These are the best representative of the session and thus have very good changes to predict correct concepts, being used together for concept prediction.  
\end{itemize}
\end{itemize}

\subsection{Discussion}
\label{discussion}

The results of our experiments enable us to positively answer our research question:

{\em H1: Can the data about the concepts frequently searched together by people within a search session be exploited to help the user explore the portions of an information space relevant to her/his information needs?}

The results show that our concept suggestion strategies have high recall, i.e., they are able to suggest most of the concepts that the users explore in the AOL-reduced dataset. However, their accuracy differs:
\begin{itemize} 
\item 
The SLACK-selective strategy is the best one from the viewpoint of precision, at least when the first queries of a session are analyzed in order to suggest concepts for query expansion.
\item 
The SLACK and the STRICT ones achieve the best recall. 
\end{itemize}
Looking at these findings, we believe that the decision of which strategy best suits concept suggestion depends on the way how we want to implement this function:
\begin{itemize}
\item
If we can suggest a larger number of concepts, then the SLACK strategy is the best one, because (i) it achieves better recall and (ii) the mean number of suggested concepts is just a bit higher than that of the other two.
\item
Differently, if we aim at precision, SLACK-selective is the preferred one at the beginning of a search session, while STRICT is more suitable for concept suggestion after the first queries.
\end{itemize}
For our current work, SLACK is the most convenient strategy because it is fairly precise and, at the same time, it suggests more relevant concepts than the other ones. Specifically, we aim at providing the user with an overview of the sets of concepts to choose from, given the search context provided by the previous queries. As the user will be enabled to choose relevant concepts in a multiple-choice box, the presence of very few irrelevant concepts, as is the case of SLACK, is not a problem.

It should be noticed that other strategies might be considered. For instance, the past search history of the individual user might be analyzed to identify the concepts he or she most frequently searched for, assuming that these represent interesting types of information. Then, this might be used to select candidate clusters for query expansion. In this paper, we left these strategies apart for two main reasons:
\begin{itemize}
\item 
Firstly, as discussed by Greenstein et al. in \cite{Greenstein-etal:17}, session-based recommendation is of primary importance because users frequently interact with systems in an anonymous way, or there might be privacy or technical reasons for avoiding the tracking of their identities. Therefore, we decided to focus on this function before exploring other scenarios.
\item
Secondly, the introduction of historic information about user behavior deserves a separate research; e.g., see \cite{White-etal:10,Sontag-etal:12,Smyth-etal:05}. Specifically, \cite{Bennett-etal:12} showed that historic information about the search behavior of an individual user helps re-ranking of search results at the beginning of a search session, while local information about the session is more effective later on, because it provides information about the actual context of search. As we deal with concept suggestion, these results cannot be immediately transferred to our case. We will thus carry out this type of analysis in a broader perspective of ontology-based personalized search support and ontology-based user models; e.g., see \cite{Teevan-etal:05,Jiang-Tan:09,Sieg-etal:10b,Gauch-etal:03,Liu-etal:02,Leung-Lee:10}.
\end{itemize}

\section{Conclusion}
\label{conclusion}

We described a session-based concept suggestion model that supports information search by proposing concepts for query expansion, given the context provided by the observed portions of the search sessions. We focus on presenting types of information that the user might also be interested in, given the data he or she is searching for, at the conceptual level.
We aim at supporting the exploration of information spaces including rather different types of data, such as those managed by Geographical Information Systems. 

Our model is based on an analysis of search behavior, that we carried out by studying the AOL query log. Specifically, our model is based on (i) the identification of clusters of concepts that frequently co-occur in search sessions, and (ii) the definition of strategies that select the clusters for concept suggestion by taking the search context that emerges from the first queries of the sessions.
The evaluation of our model has provided satisfactory results about its prediction accuracy.

Having performed off-line experiments, we could not test the serendipity of the concept suggestions, as we relied on data about the types of information that users autonomously explored in the search sessions.
Our future work includes (i) analyzing users' past behavior for improving concept suggestion, and (ii) studying the impact of our model on the serendipity of suggestions by testing it in an online experiment with users.

\section{Acknowledgments}
We thank the anonymous reviewers for their fruitful comments and suggestions. This work is partially funded by project MIMOSA (MultIModal Ontology-driven query system for the heterogeneous data of a SmArtcity, ``Progetto di Ateneo Torino\_call2014\_L2\_157'', 2015-17) and by projects "Ricerca Locale" and "Ricerca Autofinanziata" of the University of Turin. 

\balance{}

\bibliographystyle{SIGCHI-Reference-Format}


\begin{thebibliography}{00}


\ifx \showCODEN    \undefined \def \showCODEN     #1{\unskip}     \fi
\ifx \showDOI      \undefined \def \showDOI       #1{{\tt DOI:}\penalty0{#1}\ }
  \fi
\ifx \showISBNx    \undefined \def \showISBNx     #1{\unskip}     \fi
\ifx \showISBNxiii \undefined \def \showISBNxiii  #1{\unskip}     \fi
\ifx \showISSN     \undefined \def \showISSN      #1{\unskip}     \fi
\ifx \showLCCN     \undefined \def \showLCCN      #1{\unskip}     \fi
\ifx \shownote     \undefined \def \shownote      #1{#1}          \fi
\ifx \showarticletitle \undefined \def \showarticletitle #1{#1}   \fi
\ifx \showURL      \undefined \def \showURL       #1{#1}          \fi

\bibitem{Akaishi-etal:04}
{M. Akaishi}, {K. Satoh}, {and} {Y. Tanaka}. 2004.
\newblock {\em An Associative Information Retrieval Based on the Dependency of
  Term Co-occurrence}.
\newblock Springer Berlin Heidelberg, Berlin, Heidelberg, 195--206.
\newblock


\bibitem{Ardissono-etal:16}
{L. Ardissono}, {M. Lucenteforte}, {N. Mauro}, {A. Savoca}, {A. Voghera}, {and}
  {{L. La Riccia}}. 2016.
\newblock \showarticletitle{Exploration of Cultural Heritage Information via
  Textual Search Queries}. In {\em MobileHCI '16 Proceedings of the 18th Int.
  Conf. on Human-Computer Interaction with Mobile Devices and Services
  Adjunct}. ACM, 992--1001.
\newblock


\bibitem{Ardissono-etal:17d}
{L. Ardissono}, {M. Lucenteforte}, {N. Mauro}, {A. Savoca}, {A. Voghera}, {and}
  {{L. La Riccia}}. 2017a.
\newblock \showarticletitle{Semantic Interpretation of Search Queries for
  Personalization}. In {\em Proc. of UMAP 2017 Adjunct}. ACM, 101--102.
\newblock


\bibitem{Ardissono-etal:17b}
{L. Ardissono}, {M. Lucenteforte}, {N. Mauro}, {A. Savoca}, {A. Voghera}, {and}
  {L.~La Riccia}. 2017b.
\newblock \showarticletitle{OnToMap - Semantic Community Maps for knowledge
  sharing}. In {\em Proceedings of Hypertext 2017}. ACM, 317--318.
\newblock


\bibitem{Athukorala-etal:16}
{L. Athukorala}, {A. Medlar}, {A. Oulasvirta}, {G. Jacucci}, {and} {D.
  Glowacka}. 2016.
\newblock \showarticletitle{Beyond relevance: adapting exploration/exploitation
  in information retrieval}. In {\em Proc. of the 21st Int. Conf. on
  Intelligent User Interfaces ({IUI'16})}. ACM, Sonoma, California, USA,
  359--369.
\newblock


\bibitem{Ballatore-etal:16}
{A. Ballatore}, {W. Kuhn}, {M. Hegarty}, {and} {E. Parsons}. 2016.
\newblock \showarticletitle{Special issue introduction: Spatial approaches to
  information search}.
\newblock {\em Spatial Cognition \& Computation\/} {16}, 4 (2016), 245--254.
\newblock


\bibitem{Beitzel-etal:05}
{{S.M.} Beitzel}, {{E.C.} Jensen}, {O. Frieder}, {{D.D.} Lewis}, {A.
  Chowdhury}, {and} {A. Kolcz}. 2005.
\newblock \showarticletitle{Improving automatic query classification via
  semi-supervised learning}. In {\em Proc. of the Fifth IEEE Int. Conf. on Data
  Mining} {\em (ICDM '05)}. IEEE Computer Society, Washington, DC, USA, 42--49.
\newblock


\bibitem{Belkin:80}
{N.J. Belkin}. 1980.
\newblock \showarticletitle{Anomalous states of knowledge as a basis for
  information retrieval}.
\newblock {\em Canadian Journal of Information Science\/}  {5} (1980),
  133--143.
\newblock


\bibitem{Bennett-etal:12}
{{P.N.} Bennett}, {{R.W.} White}, {W. Chu}, {{S.T.} Dumais}, {P. Bailey}, {F.
  Borisyuk}, {and} {X. Cui}. 2012.
\newblock \showarticletitle{Modeling the impact of short- and long-term
  behavior on search personalization}. In {\em Proc. of the 35th Int. ACM SIGIR
  Conf. on Research and Development in Information Retrieval} {\em (SIGIR
  '12)}. ACM, New York, NY, USA, 185--194.
\newblock


\bibitem{Candan-etal:09}
{{K.S.} Candan}, {M. Cataldi}, {L.~Di Caro}, {{M.L.} Sapino}, {and} {C.
  Schifanella}. 2009.
\newblock \showarticletitle{{CoSeNa}: a context-based search and navigation
  system}. In {\em {MEDES '09} Int. Conf. on Management of Emergent Digital
  EcoSystems}. Chia Laguna, Italy, Art. 33.
\newblock


\bibitem{Cao-etal:08}
{H. Cao}, {D. Jiang}, {J. Pei}, {Q. He}, {Z. Liao}, {E. Chen}, {and} {H. Li}.
  2008.
\newblock \showarticletitle{Context-aware query suggestion by mining
  click-through and session data}. In {\em Proc. of the 14th {ACM SIGKDD} Int.
  Conf. on Knowledge Discovery and Data Mining} {\em (KDD '08)}. ACM, New York,
  NY, USA, 875--883.
\newblock


\bibitem{Chen-etal:08}
{Y. Chen}, {G-R Xue}, {and} {Y. Yu}. 2008.
\newblock \showarticletitle{Advertising Keyword Suggestion Based on Concept
  Hierarchy}. In {\em Proc. of the 2008 Int. Conf. on Web Search and Data
  Mining} {\em (WSDM '08)}. ACM, New York, NY, USA, 251--260.
\newblock


\bibitem{Fernandez-etal:11}
{M. Fern\'{a}ndez}, {I. Cantador}, {V. L\'{o}pez}, {D. Vallet}, {P. Castells},
  {and} {E. Motta}. 2011.
\newblock \showarticletitle{Semantically enhanced information retrieval: an
  ontology-based approach}.
\newblock {\em Web Semantics: Science, Services and Agents on the {W}orld
  {W}ide {W}eb\/} {9}, 4 (2011), 434--452.
\newblock


\bibitem{Gauch-etal:03}
{S. Gauch}, {J. Chaffee}, {and} {A. Pretschner}. 2003.
\newblock \showarticletitle{Ontology-based personalized search and browsing}.
\newblock {\em Web Intelli. and Agent Sys.\/} {1}, 3-4 (2003), 219--234.
\newblock


\bibitem{Giuliano-Jones:62}
{{V.E.} Giuliano} {and} {{P.E.} Jones}. 1962.
\newblock {\em Linear Associative Information Retrieval}.
\newblock {T}echnical {R}eport {ESD-TR-62-294}. {ASTIA}.
\newblock


\bibitem{GoogleKnowledgeGraph}
{Google}. 2017.
\newblock \showarticletitle{Knowledge Graph}.
  https://www.google.com/intl/it\_it/insidesearch/features/
  search/knowledge.html.
\newblock


\bibitem{Greenstein-etal:17}
{A. Greenstein-Messica}, {L. Rokach}, {and} {M. Friedman}. 2017.
\newblock \showarticletitle{Session-based recommendations using item
  embedding}. In {\em Proc. of the 22nd Int. Conf. on Intelligent User
  Interfaces} {\em (IUI '17)}. ACM, New York, NY, USA, 629--633.
\newblock


\bibitem{Gregory:10}
{S. Gregory}. 2010.
\newblock \showarticletitle{Finding overlapping communities in networks by
  label propagation}.
\newblock {\em New Journal of Physics\/} {12}, 103018 (2010).
\newblock


\bibitem{Grootjem-vanDerWeide:06}
{{F.A.} Grootjem} {and} {{T.P. van der Weide}}. 2006.
\newblock \showarticletitle{Conceptual query expansion}.
\newblock {\em Data \& Knowledge Engineering\/}  {56} (2006), 174--193.
\newblock


\bibitem{Hoeber-etal:05}
{O. Hoeber}, {{XD.} Yang}, {and} {Y. Yao}. 2005.
\newblock \showarticletitle{Conceptual query expansion}. In {\em Advances in
  Web Intelligence. {AWIC} 2005, Lecture Notes in Computer Science, vol 3528}.
  Lodz, Poland, 190--196.
\newblock


\bibitem{Huang-etal:03}
{C.-K. Huang}, {L.-F. Chien}, {and} {Y.-J. Oyang}. 2003.
\newblock \showarticletitle{Relevant term suggestion in interactive web search
  based on contextual in formation in query session logs}.
\newblock {\em Journal of the American Society for Information Science and
  Technology\/} {54}, 7 (2003), 638--649.
\newblock


\bibitem{Berg-Schuemie:99}
{{J. Van Den Berg}} {and} {M. Schuemie}. 1999.
\newblock \showarticletitle{Information retrieval systems using an associative
  conceptual space}. In {\em In Proc. of the 7th European Symposium on
  Artificial Neural Networks (ESANN'99)}. 351--356.
\newblock


\bibitem{Jiang-Tan:09}
{X. Jiang} {and} {{A.-H.} Tan}. 2009.
\newblock \showarticletitle{Learning and inferencing in user ontology for
  personalized Semantic Web search}.
\newblock {\em Information Sciences\/}  {179} (2009), 2794--2808.
\newblock


\bibitem{Jones-Purves:08}
{{C.B.} Jones} {and} {{R.S.} Purves}. 2008.
\newblock \showarticletitle{Geographical Information Retrieval}.
\newblock {\em Int. Journal of Geographical Information Science\/} {22}, 3
  (2008), 219--228.
\newblock


\bibitem{Joshi-Motwani:06}
{A. Joshi} {and} {R. Motwani}. 2006.
\newblock \showarticletitle{Keyword Generation for Search Engine Advertising}.
  In {\em 6th IEEE Int. Conf. on Data Mining - Workshops ({ICDMW'06})}. ACM,
  New York, NY, USA, 490--496.
\newblock


\bibitem{Leung-Lee:10}
{{K.W-T} Leung} {and} {Dik {D.L.}~Lee}. 2010.
\newblock \showarticletitle{Deriving concept-based user profiles from search
  engine logs}.
\newblock {\em IEEE Trans. on Knowl. and Data Eng.\/} {22}, 7 (July 2010),
  969--982.
\newblock


\bibitem{Liu-etal:02}
{F. Liu}, {C. Yu}, {and} {W. Meng}. 2002.
\newblock \showarticletitle{Personalized web search by mapping user query to
  categories}. In {\em Proc. {CIKM'02}}. McLean, Virginia, USA.
\newblock


\bibitem{Mandala-etal:99}
{R. Mandala}, {T. Tokunaga}, {and} {H. Tanaka}. 1999.
\newblock \showarticletitle{Combining multiple evidence from different types of
  thesaurus for query expansion}. In {\em Proc. of the 22nd Annual Int. ACM
  SIGIR Conf. on Research and Development in Information Retrieval ({SIGIR
  '99})}. ACM, New York, NY, USA, 191--197.
\newblock


\bibitem{Marchiorini:06}
{G. Marchiorini}. 2006.
\newblock \showarticletitle{Exploratory search: from finding to understanding}.
\newblock {\it Commun. {ACM}} {49}, 4 (2006), 41--46.
\newblock


\bibitem{Mauro-Ardissono:17b}
{N. Mauro} {and} {L. Ardissono}. 2017.
\newblock \showarticletitle{Concept-aware Geographic Information Retrieval}. In
  {\em Proceedings of 2017 IEEE/WIC/ACM Int. Conf. on Web Intelligence (WI)}.
  ACM, Leipzig, Germany.
\newblock


\bibitem{Medlar-etal:17}
{A. Medlar}, {J. Pyykk\"{o}}, {and} {D. Glowacka}. 2017.
\newblock \showarticletitle{Towards fine-grained adaptation of
  exploration/exploitation in Information Retrieval}. In {\em Proc. of the 22nd
  Int. Conf. on Intelligent User Interfaces} {\em (IUI '17)}. ACM, New York,
  NY, USA, 623--627.
\newblock


\bibitem{Molina-Bayarri:11}
{M. Molina} {and} {S. Bayarri}. 2011.
\newblock \showarticletitle{A multinational {SDI}-based system to facilitate
  disaster risk management in the {A}ndean community}.
\newblock {\em Computers \& Geosciences\/}  {18} (2011), 1501--1510.
\newblock


\bibitem{Oramas-etal:16}
{S. Oramas}, {{V.C.} Ostuni}, {T. {Di Noia}}, {X. Serra}, {and} {E. {Di
  Sciascio}}. 2016.
\newblock \showarticletitle{Sound and music recommendation with knowledge
  graphs}.
\newblock {\em {ACM} Transactions on Intelligent Systems and Technology\/} {8},
  2 (2016), Art. 21.
\newblock


\bibitem{Porrini-etal:14}
{R. Porrini}, {M. Palmonari}, {and} {G. Vizzari}. 2014.
\newblock \showarticletitle{Composite match autocompletion ({COMMA}): a
  semantic result-oriented autocompletion technique for e-marketplaces}.
\newblock {\em Web Intelligence and Agent Systems: an international journal\/}
  {12} (2014), 35--49.
\newblock


\bibitem{Qiu-Frei:93}
{Y. Qiu} {and} {{H.P} Frei}. 1993.
\newblock \showarticletitle{Concept based query expansion}. In {\em Proc. of
  the 16th Annual Int. ACM SIGIR Conf. on Research and Development in
  Information Retrieval} {\em (SIGIR '93)}. ACM, New York, NY, USA, 160--169.
\newblock


\bibitem{Ratzan:04}
{L. Ratzan}. 2004.
\newblock {\em Understanding information systems: what they do and why we need
  them}.
\newblock American Library Association, Chicago, USA.
\newblock


\bibitem{Rieh-Xie:06}
{{S.Y.} Rieh} {and} {{H. I} Xie}. 2006.
\newblock \showarticletitle{Analysis of multiple query reformulations on the
  web: the interactive information retrieval context}.
\newblock {\em Information processing and management\/}  {42} (2006), 751--768.
\newblock


\bibitem{Sieg-etal:10b}
{A. Sieg}, {B. Mobasher}, {and} {R. Burke}. 2010.
\newblock \showarticletitle{Ontology-based collaborative recommendation}.
\newblock {\em Computing\/} (2010).
\newblock


\bibitem{Smyth-etal:05}
{B. Smyth}, {E. Balfe}, {J. Freine}, {P. Briggs}, {M. Coyle}, {and} {O.
  Boydell}. 2005.
\newblock \showarticletitle{Exploiting query repetition and regularity in an
  adaptive community-based web search engine}.
\newblock {\em USER MODELING AND USER-ADAPTED INTERACTION - The Journal of
  Personalization Research\/} {14}, 5 (2005), 383--423.
\newblock


\bibitem{Sontag-etal:12}
{D. Sontag}, {K. Collins-Thompson}, {{P.N.} Bennett}, {{R.W.} White}, {S.
  Dumais}, {and} {B. Billerbeck}. 2012.
\newblock \showarticletitle{Probabilistic Models for Personalizing Web Search}.
  In {\em Proc. of the 5th ACM Int. Conf. on Web Search and Data Mining} {\em
  (WSDM '12)}. ACM, New York, NY, USA, 433--442.
\newblock


\bibitem{Teevan-etal:05}
{J. Teevan}, {{S.T.} Dumais}, {and} {E. Horvitz}. 2005.
\newblock \showarticletitle{Personalizing Search via Automated Analysis of
  Interests and Activities}. In {\em Proc. of the 28th Annual Int. ACM SIGIR
  Conf. on Research and Development in Information Retrieval} {\em (SIGIR
  '05)}. ACM, New York, NY, USA, 449--456.
\newblock


\bibitem{Voghera-etal:16}
{A. Voghera}, {R. Crivello}, {L.Ardissono}, {M. Lucenteforte}, {A. Savoca},
  {and} {{L. La Riccia}}. 2016.
\newblock \showarticletitle{Production of spatial representations through
  collaborative mapping. An experiment}. In {\em Proc. of 9th Int. Conf. on
  Innovation in Urban and Regional Planning ({INPUT 2016})}. 356--361.
\newblock


\bibitem{Wang-etal:17}
{Y. Wang}, {H. Huang}, {and} {C. Feng}. 2017.
\newblock \showarticletitle{Query expansion based on a feedback concept model
  for microblog retrieval}. In {\em Proc. of the 26th Int. Conf. on World Wide
  Web} {\em (WWW '17)}. International World Wide Web Conferences Steering
  Committee, Republic and Canton of Geneva, Switzerland, 559--568.
\newblock


\bibitem{Wang-etal:12}
{Y. Wang}, {H. Li}, {H. Wang}, {and} {{K.Q.} Zhu}. 2012.
\newblock \showarticletitle{Concept-based web search}.
\newblock In {\em Conceptual Modeling. ER 2012. Lecture Notes in Computer
  Science, vol 7532}, {P.~Atzeni} {and} {D.~Cheung} (Eds.). Springer, Berlin,
  Heidelberg, 449--462.
\newblock


\bibitem{White-etal:10}
{{R.W.} White}, {{P.N.} Bennett}, {and} {{S.T.} Dumais}. 2010.
\newblock \showarticletitle{Predicting short-term interests using
  activity-based search context}. In {\em Proc. of the 19th ACM Int. Conf. on
  Information and Knowledge Management} {\em (CIKM '10)}. ACM, New York, NY,
  USA, 1009--1018.
\newblock


\bibitem{White-etal:07}
{{R.W.} White}, {M. Bilenko}, {and} {S. Cucerzan}. 2007.
\newblock \showarticletitle{Studying the use of popular destinations to enhance
  web search interaction}. In {\em Proc. of the 30th Annual Int. ACM SIGIR
  Conf. on Research and Development in Information Retrieval} {\em (SIGIR
  '07)}. ACM, New York, NY, USA, 159--166.
\newblock


\bibitem{White-Roth:06}
{{R.W.} White} {and} {{R.A.} Roth}. 2006.
\newblock \showarticletitle{Supporting exploratory search}.
\newblock {\it Commun. {ACM}} {49}, 4 (2006), 36--39.
\newblock


\bibitem{WordNet}
{{WordNet}}. 2017.
\newblock \showarticletitle{{WordNet} - a lexical database for {English}}.
  https://wordnet.princeton.edu/.
\newblock


\bibitem{Wu-etal:12}
{W. Wu}, {H. Li}, {H. Wang}, {and} {{K.Q} Zhu}. 2012.
\newblock \showarticletitle{Probase: A Probabilistic Taxonomy for Text
  Understanding}. In {\em Proc. of the 2012 ACM SIGMOD Int. Conf. on Management
  of Data}. ACM, New York, NY, USA, 481--492.
\newblock


\end{thebibliography}

\end{document}